\documentclass[
               twocolumn,
               noshowpacs,          
               nopreprintnumbers,     
               aps,                 
               prd,                 
               a4paper,             
               superscriptaddress,  
               nofootinbib,         
               tightenlines,        
               floats,floatfix      
               ]{revtex4}

\usepackage{amsmath, amsfonts, amsthm, amssymb, graphicx}
\usepackage{epstopdf}
 \usepackage{slashed}
 \usepackage{graphicx,epsfig}
\usepackage{amsmath}
\usepackage {amssymb}
\usepackage[utf8]{inputenc}
\usepackage{hyperref}
\usepackage[francais]{babel}
\usepackage[T1]{fontenc}

\newcommand{\be}{\begin{eqnarray}}
\newcommand{\ee}{\end{eqnarray}}
\newcommand{\bea}{\begin{eqnarray}}

\newcommand{\eea}{\end{eqnarray}}

\begin{document}

\title{$D=10$ dyonic black holes in string inspired models}

\author{Gaston Giribet}
\affiliation{Physics Department, New York University, 726 Broadway, 10003, New York, USA}

\author{Marcela Lagos}
\affiliation{Departamento de F\'{\i}sica, Universidad de Concepci\'on, Casilla 160C, Concepci\'on, Chile}

\author{Julio Oliva}
\affiliation{Departamento de F\'{\i}sica, Universidad de Concepci\'on, Casilla 160C, Concepci\'on, Chile}

\author{Aldo Vera}
\affiliation{Departamento de F\'{\i}sica, Universidad de Concepci\'on, Casilla 160C, Concepci\'on, Chile}


\date{\today}

\begin{abstract}
We consider string inspired models in $D=10$ spacetime dimensions, which include couplings with 1- and 3-form fields as well as $R^4$ higher curvature corrections to the gravitational action. For such models, we explicitly construct a family of black hole solutions with both electric and magnetic charges, and with different horizon topologies. The solutions exhibit some features similar to those of self-gravitating monopoles in Einstein-Yang-Mills theory, which we discuss. When higher-curvature corrections are switched off, our solutions reduce to charged $p$-brane solutions previously studied in the literature. Novel qualitative features appear due to the $R^4$ terms, though. Such is the case of the emergence of branch singularities for charged solutions that, nonetheless, can be shielded by the event horizon. 
\end{abstract}

\maketitle

\section{Introduction} 

String theory induces modifications to Einstein gravity which are represented by the presence of higher-curvature terms in the low energy effective action. Type II string theories, for example, contain quartic ($R^4$) terms which appear at cubic order in the $\alpha'$ expansion. The theory also contains additional fields; in the case of type IIA strings, apart from the fields that are already present in the bosonic sector, the theory contains $1$- and $3$-form fields, together with their magnetic duals. 

The quartic modification to Einstein-Hilbert action in the Type II theories at tree-level has the form 
\begin{equation} 
\int \Big( c_1\ t_8 t_8 + c_2 \  \epsilon_{10}\cdot \epsilon_{10}\Big) R^4 \ , \label{uno}
\end{equation}
where $c_1=8c_2$ is a positive coupling, $\epsilon_{10}$ is the Levi-Civita pseudo-tensor in dimension $10$, which here appears with two of its indices contracted with those of a second copy of itself, and $t_8$ is a rank-8 tensor introduced in \cite{Gross:1986iv}; see also \cite{Giusto:2004xm}. $R^4$ represents in (\ref{uno}) for the tensor product of four Riemann tensors. This yields 
\begin{equation}
12\int \sqrt{-g} \Big( (c_1-8c_2)(R_{\mu \nu \rho \sigma}R^{\mu \nu \rho \sigma})^2 + ...\Big) , 
\end{equation}
where the ellipsis stand for other contractions of four Riemann tensors. There are, in addition, a dilaton envelope factor and couplings between the curvature and the Kalb-Ramond field, which we are omitting here. Notice that for $c_1=8c_2$ the squared Ktretschmann scalar disappears from the action, which facilitates Einstein spaces to persist as solutions. 

The condition $c_1\geq 0$ is crucial for a theory like (\ref{uno}) to avoid superluminal behavior \cite{Gruzinov:2006ie}. For simplicity, we will focus on the case $c_1=0\neq c_2$. Remarkably, for such particular choice of $R^4$ terms we will be able to write down black hole solutions charged under different $p$-form fluxes analytically. The simplification with respect to the case $c_1\neq 0$ comes from the fact that the terms $\epsilon_{10}\cdot \epsilon_{10}R^4$ correspond to the dimensional extension of the Pfaffian whose integral, in dimension 8, computes the Euler characteristic in virtue of the Chern-Weil-Gauss-Bonnet theorem. The topological origin of this term is what makes the field equations to be of second order and, thus, tractable analytically. Being of second order, the theory is, in addition, free of Ostrogradsky instabilities; the perturbative theory results free of ghosts around flat space \cite{Zwiebach:1985uq}. Besides, the theory with such $R^4$ terms, provided neither quadratic nor cubic terms are present, does not exhibit the causality problems of the type discussed in \cite{Camanho:2014apa}. All this makes the case $c_1=0$ an excellent arena to investigate the qualitative features that the presence of $R^4$ terms may introduce.

In $D=10$ dimensions we thus consider the action
\begin{equation}
I=\frac{1}{16\pi \ell_p^8} \int \Big(R- F_{(2)} \wedge *F _{(2)}- F_{(4)} \wedge *F _{(4)}+\frac{\alpha^3}{2^4} \epsilon_8 \epsilon_8 R^4\Big)
\end{equation}
where $\ell_P$ is the Planck length. $F_{(2)}$ and $F_{(4)}$ are the field strengths associated to the 1-form $A_{(1)}$ and the 3-form $A_{(3)}$, respectively ($F_{(p+1)}=dA_{(p)}$). Conventions are such that $\int F_{(p)} \wedge *F _{(p)}= (1/p)\int d^{10}x \sqrt{-g}F_{\mu_1..\mu_p} F^{\mu_1..\mu_p}$. One can also add a cosmological constant term $-1/(8\pi \ell_p^8)\int d^{10}x\sqrt{-g}\Lambda$. The coupling constant $\alpha $ has mass dimension $-2$. The tensor structure in the quartic terms is given by $\epsilon_8 \epsilon_8=(1/2)\epsilon_{10}\cdot \epsilon_{10}$, which can be conveniently written as 
\begin{equation}
\epsilon_8 \epsilon_8R^4=\delta^{\mu_1 \mu_2 ... \mu_8}_{\nu_1 \nu_2 ... \nu_8}R^{\nu_1 \nu_2}_{\ \ \ \ \mu_1 \mu_2}R^{\nu_3 \nu_4}_{\ \ \ \ \mu_3 \mu_4}R^{\nu_5 \nu_6}_{\ \ \ \ \mu_5 \mu_6}R^{\nu_7 \nu_8}_{\ \ \ \ \mu_7 \mu_8},
\end{equation}
where $\delta$ is the totally anti-symmetric Kronecker symbol \cite{Lovelock:1971yv}. The sign $\alpha^3>0$ is the one that yields a consistent model; see \cite{Gruzinov:2006ie}. 

For the theory above, we will write down analytic dyonic black hole solutions charged under both the $1$- and $3$-form fields, and with different horizon topologies. The local geometry of the base manifolds (i.e. of the constant-time sections of the horizons) of these solutions will be the direct product of $2^m$ copies $(2^{3-m})$-dimensional constant-curvature manifolds, with $m=\{ 0,1,2,3 \}$. For example, in the case of such constant-curvature manifolds being $(2^{3-m})$-spheres, the case $m=0$ would correspond to the $10$-dimensional Reissner-Nordstr\"{o}m black hole, whose base manifold is a $8$-sphere. The other extreme of the list, the case $m=3$, would correspond to a topological black hole with flat horizon; namely a black brane. Higher-genus topological black holes with compact horizons require to consider as base manifold the product of Fuchsian quotients of hyperbolic spaces. The two most interesting and less simple examples are black hole solutions whose horizons are either products of four $2$-spheres or of two $4$-spheres. These solutions are supported by magnetic fluxes on the spheres and admit net electric charges. The latter are the cases we will study in this paper.

\section{Black hole solutions}

\subsection{The solution with $S^2\times S^2 \times S^2 \times S^2$}

Let us start with the case $m=2$; that is, solutions whose base manifolds are direct product of four constant-curvature 2-manifolds. We begin by considering $2$-spheres. The configuration of the 3-form for such a solution is given by
\begin{equation}
F_{(4)} = \frac{Q_2}{r^4}dt\wedge dr \wedge \sum_{i=1}^4 \text{vol}(S^2_i) + P_2 \sum_{i<j}^4 \text{vol}(S^2_i)\wedge \text{vol}(S^2_j) 
\end{equation}
where $\text{vol}(S^2_i)$ stands for the volume form of the $i^{\text{th}}$ 2-sphere $S^2_i$, $i=\{1,2,3,4\}$. $Q_2$ and $P_2$ are two integration constants associated to the electric and magnetic charges under the 3-form field $A_{(3)}$. Similarly, the configuration of the 1-form is given by 
\begin{equation}
F_{(2)} = \frac{Q_0}{r^8}dt\wedge dr  
\end{equation}
which correspond to the Coulombian potential $A_{(1)}=Q_0/(7r^{7})dt$. 

The metric takes the form
\begin{equation}
ds^2 = - H(r) \ dt^2 +\frac{dr^2}{H(r)} + r^2 \sum_{i=1}^4 \frac{dz_i d\bar{z}_i}{(1+\frac{\sigma}{4}z_i\bar{z}_i)^2}
\end{equation}
where $(z_i, \bar{z}_i)$ are complex projective variables on the $i^{\text{th}}$ 2-(pseudo)sphere. $\sigma =1$ corresponds to the metric of the base manifold to be that of a product of four unit 2-spheres. The cases $\sigma=0,-1$ correspond to the locally flat spaces and locally hyperbolic spaces, respectively. $H$ is a function of the radial coordinate $r$, which is given by the fourth-order polynomial equation 
\begin{eqnarray}
&&\alpha^3 r^8 (g_0 H^4-g_1\sigma H^3+g_2 |\sigma |H^2 -g_3\sigma H+g_4|\sigma |)-\nonumber \\
&&g_5r^{14}H=T \label{Poli1}
\end{eqnarray}
with the matter content given by
\begin{eqnarray}
T=-t_1Q_0^2-t_2Q_2^2r^4+t_3\ell_P^8Mr^7+t_4P_2^2r^8-t_5\sigma r^{14} \nonumber
\end{eqnarray}
where the coefficients $g_i$ and $t_i$ in their prime factorization forms are
\begin{eqnarray}
g_0 &=&2^7 \cdot 3^4 \cdot 5^2 \cdot 7^2 , \  g_1 =2^9 \cdot 3^4 \cdot 5^2 \cdot 7 , \nonumber \\
g_2 &=&2^8 \cdot 3^5 \cdot 5 \cdot 7  , \ \ \ g_3 = 2^9 \cdot 3^3 \cdot 5 \cdot 7, \nonumber \\
g_4 &=& 2^7 \cdot 3^3 \cdot 5 \cdot 7 , \ \ \ g_5 = 2^3 \cdot 3^2 \cdot 5 \cdot 7 . \nonumber 
\end{eqnarray} 
and
\begin{eqnarray}
t_1 &=& 3^2\cdot 5, \ \ t_2 = 2^3\cdot 3^2\cdot 5\cdot 7, \nonumber \\
t_3 &=& 2^5\cdot 3^2\cdot 5\cdot 7 \ \pi , \ \ t_4 = 2^2\cdot 3^4\cdot 5 \cdot 7 , \nonumber \\
t_5 &=&  2^3\cdot 3^2\cdot 5. \nonumber 
\end{eqnarray}

If $(\hat{\lambda} \alpha /64\pi \ell_P^8) \int d^{10}x\sqrt{-g} \delta^{\mu_1 \mu_2 \mu_3 \mu_4}_{\nu_1 \nu_2 \nu_3 \nu_4}R^{\nu_1 \nu_2}_{\ \ \ \ \mu_1 \mu_2}R^{\nu_3 \nu_4}_{\ \ \ \ \mu_3 \mu_4}$ is also included in the action, then the left hand side of (\ref{Poli1}) receives a piece $\hat{\lambda } \alpha r^{12} (g_6 H^2-g_7\sigma H+g_8|\sigma |)$ with $g_6=2^4\cdot 3^3 \cdot 5 \cdot 7^2$, $g_7=2^5 \cdot 3^3\cdot 5 \cdot 7$, $g_8=2^4\cdot 3^3\cdot 7$. 

In the case of non-vanishing cosmological constant ($\Lambda \neq 0$), the matter content $T$ receives an additional term $70 \Lambda r^{16}$. The general relativity (GR) limit (or, equivalently, the large $r^2/\alpha$ limit) of the solution yields
\begin{equation}
H\simeq -\frac{r^2\Lambda}{36}+\frac{\sigma}{7}-\frac{4\pi\ell_P^8 M}{r^7}-\frac{9P_2^2}{2r^6}+\frac{Q_2^2}{r^{10}}+\frac{Q_0^2}{56r^{14}}+ ...\label{nueve}
\end{equation}
where the ellipsis stand for subleading terms in orders of $\alpha /r^2$. This is consistent with the results for charged $p$-brane solutions with non-spherical horizons in Einstein theory\footnote{We thank Marcello Ortaggio for calling our attention to references \cite{Bardoux:2012aw, Bardoux:2010sq} which contains black hole solutions to which ours tend in the limit $\alpha^3\to 0$.}. Solutions with base manifolds that are not constant-curvature manifolds where studied, for instance, in references \cite{Gibbons:2002pq, Dotti:2005rc}; see also references thereof.

Notice that, in (\ref{nueve}), the term with the charge $P_2$ presents a damping-off weaker than the Newtonian potential in $10$ dimensions, what typically leads to a divergent gravitational energy. This is reminiscent of what happens with self-gravitating monopoles in Einstein-Yang-Mills theory \cite{Gibbons:2006wd}. In addition, the sign of such energy contribution to the gravitational potential seems to introduce instabilities.

The electrically charged solution for finite $\alpha $ also presents curious features. In particular, there exist a branch singularity that occurs at a finite distance that, in principle, can be smaller that the horizon location $r_+$. Such branch singularity happens when radicands in the solution to a polynomial equation such as (\ref{Poli1}) take negative values. This happens because of the relative signs between the Newtonian term and the terms with the charges $Q_{0,2}$, what makes radicands to vanish for finite $r$. This results in non-real components of the metric and divergences in the curvature invariants, what are typical features of charged solutions in Lovelock theory.

\begin{figure}[h]
\includegraphics[scale=0.3]{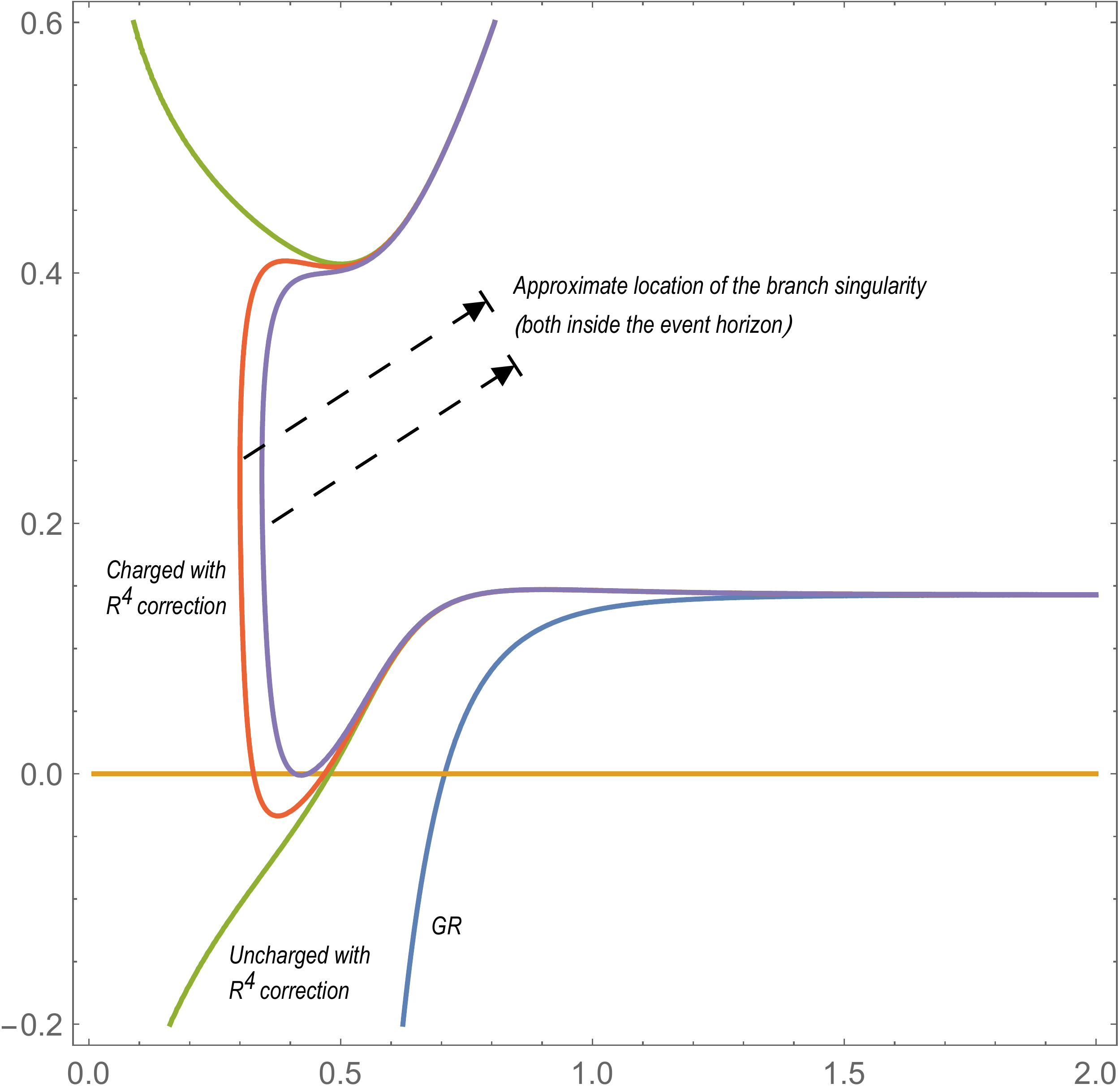}
\caption{Lapse functions for the $(S^2)^4$ charged under a Maxwell field and uncharged under $F_{(4)}$. We have used $M=10^{-3}$ and the doublets $(Q_0,\alpha)$ equal to $(0,0)$-blue, $(0,8\times 10^{-2})$-green, $(10^{-2},8\times 10^{-2})$-red and $(1.55\times 10^{-2}, 8\times 10^{-2})$-purple.}
\end{figure}
\begin{figure}[h]
\includegraphics[scale=0.3]{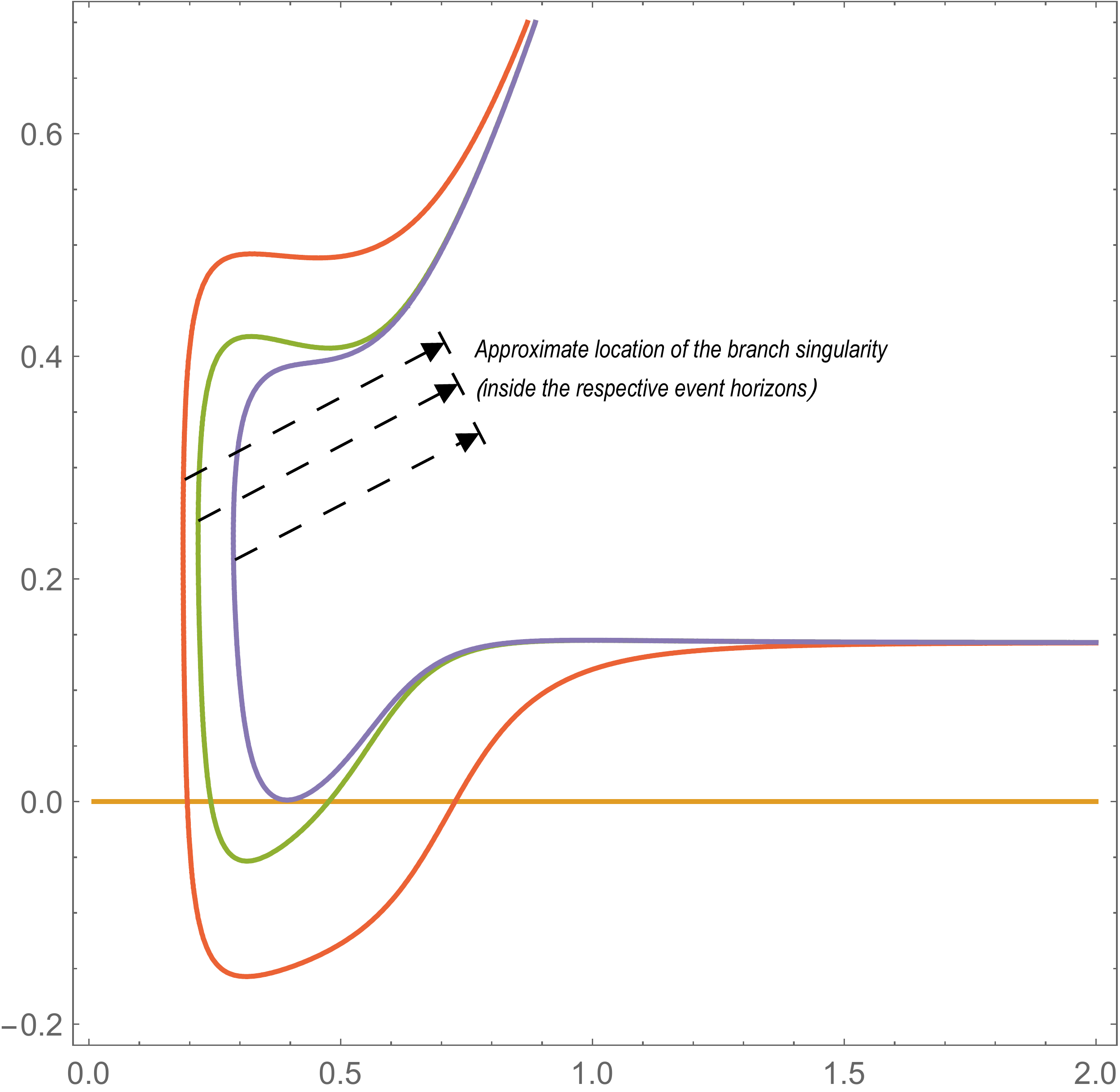}
\caption{Lapse functions for the $(S^2)^4$ dyonic-$F_{(4)}$ black holes and uncharged under Maxwell. We have used $\alpha=0.08$, $Q_0=0$ and $M=10^{-3}$ as well as the doublets $(Q_2,P_2)$ with $(1.45\times 10^{-2},2\times 10^{-2})$-purple, $(10^{-2},2\times 10^{-2})$-green and $(10^{-2},8\times 10^{-2})$-red.}
\end{figure}
It is worth mentioning that, if we add to the action a term $(\lambda\alpha /64\pi\ell_P^8)\int d^{10}x \sqrt{-g}(F_{\mu_1 \mu_2 \mu_3 \mu_4} F^{\mu_1 \mu_2 \mu_3 \mu_4})^2$ and consider the case $P_2=Q_0=0$, then equation (\ref{Poli1}) receives the following contribution on the right hand side
\begin{equation}
2^3\cdot 3^3\cdot 5\cdot 7\ r^7\int^r_{r_0}dr\ X^2(r)(2^53^2\alpha \lambda +r^4\lambda)
\end{equation}
where $X(r)$ is solution to the equation 
\begin{equation}
X(r)r^4+2^63\alpha \lambda X^3(r)=Q_2 . 
\end{equation}
The ultraviolet cut-off $r_0$ above contributes to the mass.

\subsection{The solution with $S^4 \times S^4$}

Solutions of a similar type exist for base manifolds that are the direct product of two constant-curvature 4-manifolds. This corresponds to $m=1$. In this case, the magnetic configuration of the 3-form is given by
\begin{equation}
F_{(4)} =\hat{P}_2 \sum_{i=1}^2 \text{vol}(S^4_i) \ , \ \ \ \text{vol}(S^4_i)=\frac{dz_i\wedge d\bar{z}_i \wedge dw_i\wedge d\bar{w}_i}{(1+\frac{1 }{4}z_i\bar{z}_i+\frac{ 1}{4}w_i\bar{w}_i)^{4}}
\end{equation}
where $\text{vol}(S^4_i)$ stands for the volume form of the 4-sphere $S^4_i$, $i=\{1,2\}$. Again, the integration constant $\hat{P}_2$ is associated to the magnetic charge. The metric for this solution (with $\sigma =1$) takes the form
\begin{equation}
ds^2 = - \hat{H}(r)\ dt^2 +\frac{dr^2}{\hat{H}(r)} + r^2 \sum_{i=1}^2 \frac{dz_i d\bar{z}_i+dw_i d\bar{w}_i}{(1+\frac{1}{4}z_i\bar{z}_i+\frac{1}{4}w_i\bar{w}_i)^2}.
\end{equation}
We have that the local geometry of the base manifold is now given by a product of two unit 4-spheres (or, more generally, of a pair of identical 4-manifolds of constant-curvature $\sigma =\{ 0,\pm 1 \}$). For this case, $\hat{H}$ is given by the following fourth-order polynomial equation 
\begin{eqnarray}
&&\alpha^3 r (\hat{g}_0\hat{H}^4 -\hat{g}_1 \sigma \hat{H}^3 + \hat{g}_2 |\sigma | \hat{H}^2 - \hat{g}_3 \sigma \hat{H}+\hat{g}_4|\sigma |)+ \nonumber \\
&&\ (-\hat{g}_5 \hat{H}+\hat{g}_6\sigma ) r^7 =T \label{Poli2}
\end{eqnarray} 
with the matter contribution
\begin{eqnarray}
T=\hat{t}_1\ell_P^8M+\hat{t}_2\hat{P}_2^2r+\hat{t}_3\Lambda r^9,
\end{eqnarray} 
where the coefficients $\hat{g}_i$ are given by
\begin{eqnarray}
\hat{g}_0 &=&2^6\cdot 3^4\cdot 5^2\cdot 7^2 , \  \hat{g}_1 = 2^8\cdot 3^5\cdot 5^2\cdot 7, \nonumber \\
\hat{g}_2 &=& 2^7\cdot 3^5\cdot 5\cdot 7^2 , \ \hat{g}_3 = 2^8\cdot 3^5 \cdot  5 \cdot  7 \nonumber \\
\hat{g}_4 &=& 2^6\cdot  3^5 \cdot  5\cdot  7, \ \hat{g}_5 = 2^2\cdot 3^2\cdot 5\cdot 7 \nonumber \\
\hat{g}_6 &=& 2^2\cdot 3^3\cdot 5 , \nonumber 
\end{eqnarray} 
and the coefficients $\hat{t}_i$ are given by
\begin{eqnarray}
\hat{t}_1 = 2^4\cdot 3^2\cdot 5\cdot 7\ \pi , \ \ \hat{t}_2 = 2\cdot 3^3\cdot 5\cdot 7 , \hat{t}_3 =5\cdot 7 \nonumber 
\end{eqnarray} 

The GR limit of this solution yields
\begin{equation}
\hat{H}\simeq -\frac{\Lambda r^2}{36} +\frac{3\sigma }{7}-\frac{3\hat{P}_2^2}{2r^6}-\frac{4\pi \ell_P^8M}{r^7}+ ...\label{a16}
\end{equation}
where the ellipsis stand for subleading terms in orders of $\alpha /r^2$. Again, we observe the weakened damping-off of the term with charge $\hat{P}_2$.  The fact that GR solution (\ref{a16}) is recovered when $\alpha= 0$ shows that we are considering a perturbative solution of the higher-curvature theory. These theories typically present up to four different branches of static solutions, some of them non-existent (for divergent) in the $\alpha\to 0$ limit.

If $(\hat{\lambda} \alpha /64\pi \ell_P^8) \int d^{10}x\sqrt{-g} \delta^{\mu_1 \mu_2 \mu_3 \mu_4}_{\nu_1 \nu_2 \nu_3 \nu_4}R^{\nu_1 \nu_2}_{\ \ \ \ \mu_1 \mu_2}R^{\nu_3 \nu_4}_{\ \ \ \ \mu_3 \mu_4}$ is included in the action, then the left hand side of (\ref{Poli2}) receives a piece $\hat{\lambda } \alpha r^5(\hat{g}_7 |\sigma |-\hat{g}_8\sigma \hat{H}+\hat{g}_9\hat{H}^2)$ with $\hat{g}_7= 2^3\cdot 3^3\cdot 7^2$, $\hat{g}_8=2^4\cdot 3^4\cdot 5\cdot 7$, $\hat{g}_9=2^3\cdot 3^3\cdot 5\cdot 7^2$, and it also leads to a solvable equation.

\subsection{Thermodynamics}

The qualitative new features introduced by the higher-curvature terms in the action can also be analyzed by taking a glance at the thermodynamics of the solution. For instance, consider the black hole solution with $m=1$ and $P_2=0=\Lambda$, whose Hawking temperature, including both $R^2$ and $R^4$ terms, is given by
\begin{equation}
T_{\text{H}}=\frac{8\sigma r_+^{14}+48\alpha \hat{\lambda} |\sigma |r_+^{12}+384\alpha^3|\sigma |r_+^8-24 Q_2^2r_+^4 -Q_0^2}{32\pi r_+^9(192 \alpha^3\sigma +12\alpha\hat{\lambda }\sigma r_+^4+r_+^6)}.\label{l17}
\end{equation}
We see from this the short-distance modifications to the GR behavior. At large $r_+^2/\alpha$, (\ref{l17}) clearly reproduces the thermodynamical behavior of asymptotically flat black hole solutions to Einstein equations. In contrast, provided $\alpha\neq 0$, the small $r_+$ behavior deviates from that of GR.
  
Short distance contributions are also seen in the black hole entropy formula; namely
\begin{equation}
S_{\text{BH}}=\frac{(4\pi)^4r_+^8}{4\ell_P^8} + \frac{4\sigma \hat{\lambda }\alpha r_+^6}{\ell_P^8}+\frac{192(4\pi)^4\alpha^3\sigma r_+^2}{\ell_P^8},
\end{equation}
from which we observe that the Bekenstein-Hawking area-law term gets supplemented by contributions of higher-curvature terms that vanish in the limit $\alpha\to 0$. Such terms scale with lower powers of the volume, as expected both from dimensional analysis. Unlike the term proportional to the horizon area, the sign of the other two terms is sensitive to the curvature of the base manifold ($\sigma $) as well as to the sign of the coupling constant of the higher-curvature terms ($\alpha $). Different choices of those signs lead to different qualitative behaviors, including some pathological ones. Such a dependence on the signs of the couplings is also observed in the expression for the temperature, although, as expected, becomes relevant only at short distances.    

\section{Conclusions}

For string inspired models that include $R^4$ terms together with $F_{(2)}$ and $F_{(4)}$ fluxes, we have derived analytic dyonic black hole solutions, with different horizon topologies. We focus our attention to those 10-dimensional solutions whose base manifolds were given by the tensor product of constant-curvature manifolds. We have shown that magnetically charged solutions of this type exhibit weakened asymptotics that are reminiscent of those of self-gravitating monopoles in Einstein-Yang-Mills theory. We have also shown that the higher-curvature terms make the electrically charged solutions to develop a branch singularity at a finite distance, which can well be shielded by the event horizon. It is remarkable that, even in the case in which both $R^2$ and $F_{(4)}^4$ terms are included in the action, the particular Lovelock-type $R^4$ tensor structure we considered here allowed us to write down dyonic black hole solutions in a closed form. This provides a tractable model to study the physics of higher-curvature effects analytically.

\[ \]
The authors thank Marcello Ortaggio for discussions. The work of G.G. is supported by the NSF through grant PHY-1214302.
M.L. and A.V. appreciate the support of CONICYT Fellowships No. 21141229 and No. 21151067, respectively. J.O. is supported by CONICYT grant 1181047.
\[ \]

\providecommand{\href}[2]{#2}\begingroup\raggedright\endgroup
\end{document}